\documentclass{article}
\usepackage{spconf,amsmath,graphicx}

\usepackage{enumitem}
\setlist{nosep, leftmargin=14pt}

\usepackage{mwe} 
\usepackage{xcolor}
\usepackage{multirow}
\usepackage{amssymb}
\usepackage{dblfloatfix}

\title{Incorporating Boundary Uncertainty into Loss Functions for Biomedical Image Segmentation}
\makeatletter
\def\@name{ \emph{Michael Yeung}$^{1,2,3}$, \emph{Guang Yang}$^{3}$, \emph{Evis Sala}$^{2,4}$, \emph{Carola-Bibiane Sch\"{o}nlieb}$^{5}$, \emph{Leonardo Rundo}$^{2,4}$\\}
\makeatother

\address{$^1$ School of Clinical Medicine, University of Cambridge, Cambridge, UK\\
$^2$ Department of Radiology, University of Cambridge, Cambridge, UK\\
$^3$ National Heart \& Lung Institute, Imperial College London, London, UK \\
$^4$ Cancer Research UK Cambridge Centre, University of Cambridge, Cambridge, UK\\
$^5$ Department of Applied Mathematics and Theoretical Physics, University of Cambridge, Cambridge, UK\\
}
\begin{document}
%
\maketitle
\begin{abstract}
Manual segmentation is used as the gold-standard for evaluating neural networks on automated image segmentation tasks. Due to considerable heterogeneity in shapes, colours and textures, demarcating object boundaries is particularly difficult in biomedical images, resulting in significant inter and intra-rater variability. Approaches, such as soft labelling and distance penalty term, apply a global transformation to the ground truth, redefining the loss function with respect to uncertainty. However, global operations are computationally expensive, and neither approach accurately reflects the uncertainty underlying manual annotation. In this paper, we propose the Boundary Uncertainty, which uses morphological operations to restrict soft labelling to object boundaries, providing an appropriate representation of uncertainty in ground truth labels, and may be adapted to enable robust model training where systematic manual segmentation errors are present. We incorporate Boundary Uncertainty with the Dice loss, achieving consistently improved performance across three well-validated biomedical imaging datasets compared to soft labelling and distance-weighted penalty. Boundary Uncertainty not only more accurately reflects the segmentation process, but it is also efficient, robust to segmentation errors and exhibits better generalisation. 

\end{abstract}


\section{Introduction}

Manual segmentation of biomedical images is performed by humans, often requiring expert knowledge, to assign each pixel of an image to a class. Not only is this process time-consuming, but the difficult, and often subjective nature, results in significant inter and intra-rater variability \cite{warfield2004simultaneous}. Automatic methods for biomedical image segmentation aim to address these issues, using manual segmentations to provide labels for supervised training and performance evaluations. However, categorical labels from manual segmentation fail to capture the uncertainty associated with the process of humans assigning class labels.

A common approach in classification tasks to represent uncertainty is to apply soft labels (SLs) to the ground truth, converting binary decisions into probabilistic scores \cite{nguyen2014learning, szegedy2016rethinking}. Segmentation describes a pixel-wise classification task, and the equivalent therefore involves assigning soft labels to each pixel \cite{wang2020improved}. However, this form of global uncertainty ignores important spatial information, and is different from uncertainty in manual segmentation which is instead concentrated around class boundaries \cite{krygier2021quantifying}. 

To focus optimisation on boundaries, the distance penalty term (DPT) computes Distance Transform Maps (DTMs) based on Euclidean distances to penalise predictions relative to class boundaries \cite{kervadec2019boundary, sugino2021loss}. However, generating DTMs are not only computationally expensive, but also increases the risk of overfitting by focusing too strictly on classifying boundary pixels according to the ground truth despite the uncertainty.

The main contributions of this work can be summarised as follows:
\begin{enumerate}
\item We propose a new ground truth transformation, known as Boundary Uncertainty (BU), which uses morphological operations to restrict soft labelling to boundary regions, providing an efficient approximation to segmentation uncertainty.
\item We demonstrate consistently improved performance with the balanced variant of BU across three well-validated datasets over SL and DPT.
\item We show the unbalanced variant of BU enables robust training when dealing with over-segmented and under-segmented labels. 
\end{enumerate}

\section{Materials and methods}

Boundary Uncertainty combines soft labelling with morphological operations to approximate segmentation uncertainty. Here, we first introduce soft labelling, followed by morphological operations, before describing Boundary Uncertainty, and conclude with dataset and implementation details. For all experiments, we use the original 2D U-Net as our baseline model, and apply ground truth transformations to the Dice loss \cite{ronneberger2015u, milletari2016v, yeung2021unified}. 

\subsection{Soft labels}

Manual segmentation involves assigning class labels to each pixel in an image. In a binary segmentation task using one-hot encoded softmax outputs, the value for a pixel, $p$, at location $x$ in the ground truth is defined as:

\begin{equation}
p_{x}: \begin{cases}
p = 1, & \text { if } x \in \mathcal{T} \\
p = 0, & \text { if } x \notin \mathcal{T}
\end{cases},
\label{eq:predProb}
\end{equation}
where $\mathcal{T}$ is the segmentation target. 

To represent uncertainty, SL involves converting class labels into probabilistic scores:

\begin{equation}
p_{x}: \begin{cases}
p \leqslant 1, & \text { if } x \in \mathcal{T} \\
p \geqslant 0, & \text { if } x \notin \mathcal{T}
\end{cases},
\label{eq:predProb}
\end{equation}
subject to:

\begin{equation}
p_{x \notin \mathcal{T}} \leqslant p_{x \in \mathcal{T}}.
\label{eq:predProb}
\end{equation}

Greater confidence in classifying a pixel as the segmentation target or background is represented as values closer to 1 and 0, respectively. To approximate manual segmentation uncertainty, we restrict SL to boundary regions using morphological operations. 

\subsection{Morphological operations}

Morphological operations involve non-linear, local transformations on region boundaries. The two basic morphological transformations are erosion and dilation, applied to binary images by querying with a structuring element at all positions. Let $I$ represent the input grey-scale image of size $m \times n$. Dilation ($\oplus$) and erosion ($\ominus$) operations are defined as:

\begin{equation}
\left(I \oplus W\right)(x, y)=\max _{i \in S_{1} j \in S_{2}}\left(I(x-i, y-j)+W(i, j)\right),
\end{equation}

\begin{equation}
\left(I \ominus W\right)(x, y)=\min _{i \in S_{1} j \in S_{2}}\left(I(x+i, y+j)-W(i, j)\right),
\end{equation}
where $W$ is the structuring element. 

\subsection{Boundary Uncertainty}

\begin{figure}[ht!]
    \centering
    \includegraphics[scale=0.35]{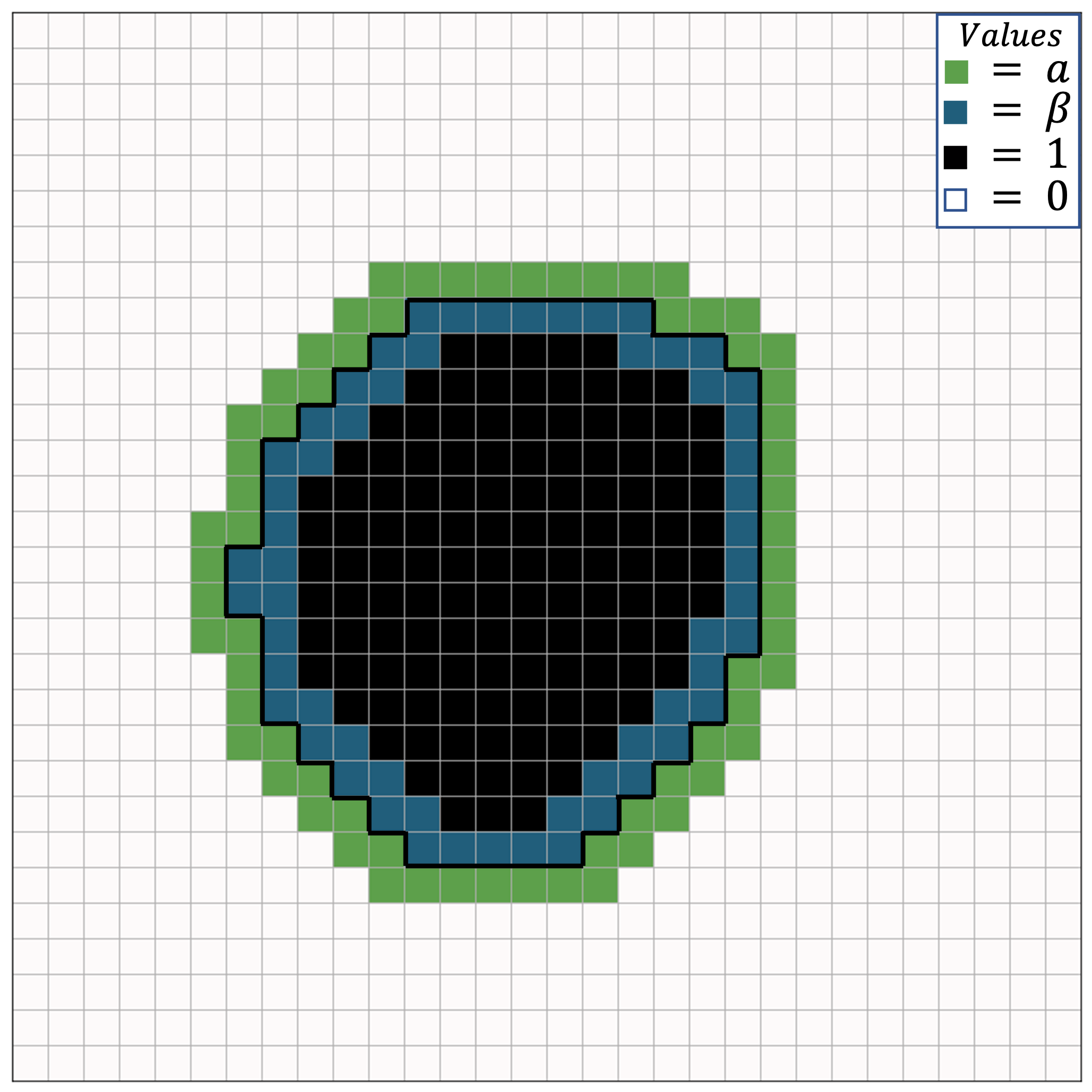}
    \caption{Example of applying BU to a manual segmentation. The black border delineates the original manual segmentation. Morphological dilation expands the segmentation target into the green region, where values are assigned $\alpha$. In contrast, morphological erosion subtracts the blue region, where values are assigned $\beta$. Due to the local nature of morphological operations, all other pixels remain unaffected.}
    \label{fig:figure_1}
\end{figure}

BU assigns pixels with probabilistic scores around boundary regions:

\begin{equation}
p_{x \in \mathcal{T}}: \begin{cases}
p = \alpha, & \text { if } p \in ((I \oplus W)_{n_\textbf{iter}} - I) \\
p = \beta, & \text { if } p \in (I - (I \ominus W)_{n_\textbf{iter}})
\end{cases},
\label{eq:predProb}
\end{equation}
where $n_\textbf{iter}$ denotes the number of iterations of the morphological operation, and $\alpha$ and $\beta$ are hyperparameters that determine the SL values directly exterior and interior to the boundary region respectively (Fig.~\ref{fig:figure_1}). The balanced form is subject to the constraint that $\alpha + \beta = 1$ and $\alpha \geqslant \beta$, used for when segmentation errors are equally likely to occur either side of the boundary. The setting $\alpha = 1$ and  $\beta = 0$ corresponds to the original hard labels.

In contrast, the unbalanced form restricts $0 \leqslant \alpha + \beta \leqslant 2$ and $\alpha \geqslant \beta$, used when there is systematic bias towards under-estimation or over-estimation of the segmentation target. The unbalanced form has been applied in highly class imbalanced segmentation, where dilation of the segmentation target improves recall with small segmentation targets \cite{du2019boundary, wang2018focal}. 

$n_\textbf{iter}$ determines the extent of BU, with larger values of $n$ corresponding to larger regions of uncertainty.

\begin{figure}[ht!]
    \centering
    \includegraphics[scale=0.35]{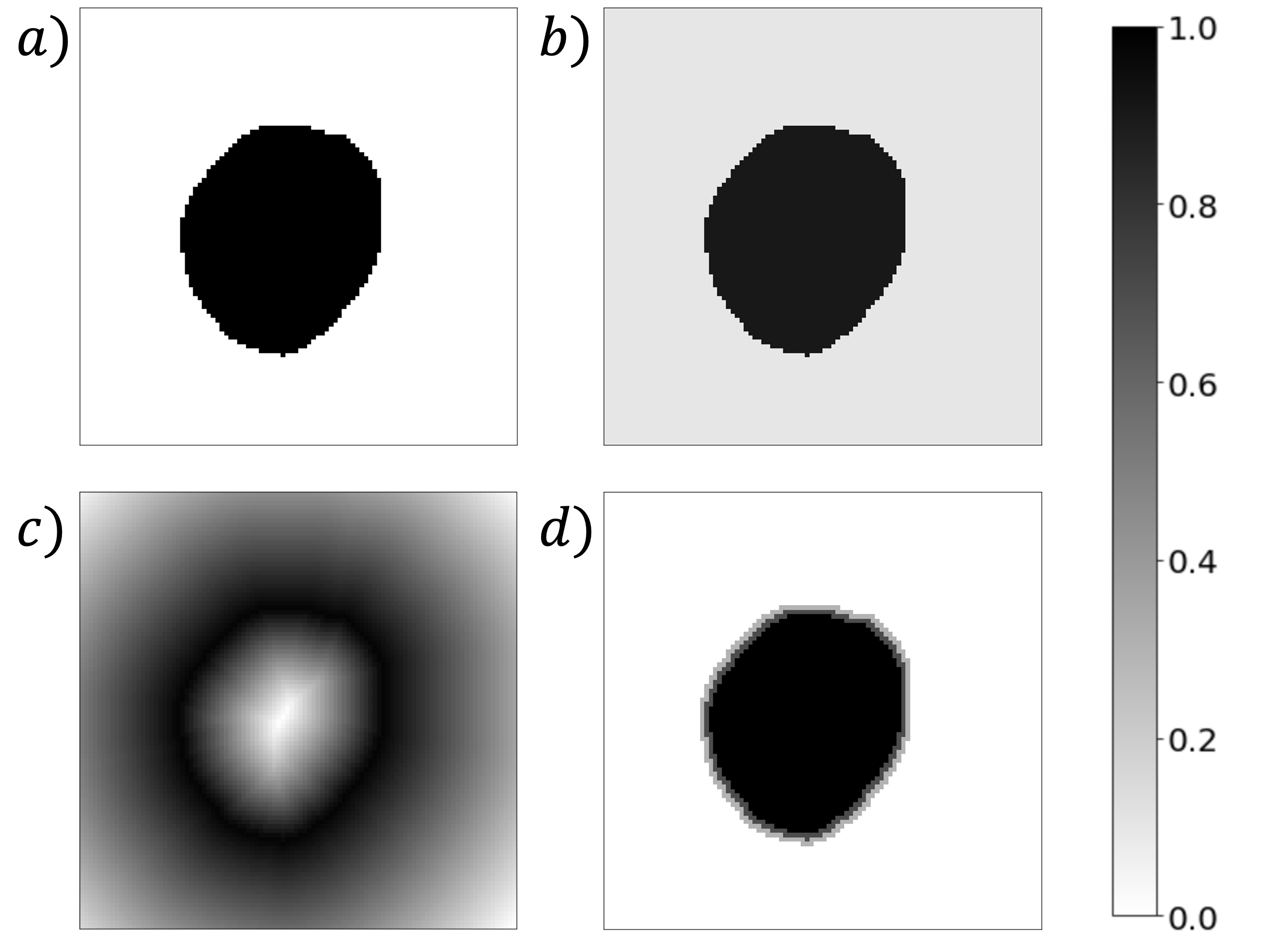}
    \caption{Ground truth transformations where a) ground truth, b) SL, c) DPT and d) BU with $\alpha = 0.7, \beta = 0.3$ and $n = 1$.}
    \label{fig:figure_2}
\end{figure}

In summary, BU enables robust model training by approximating segmentation error, which is mainly concentrated around boundaries, regardless of whether the errors are random or systematic in origin. The effect of different ground truth transformations are shown in Fig.~\ref{fig:figure_2}.

\begin{table*}[t]
\centering
\caption{Hyperparameter tuning of BU on the DRIVE, 2018DSB and CVC-ClinicDB datasets. The highest scores are denoted in bold.}
\scalebox{0.75}{
\begin{tabular}{ll|lll|lll|lll}
\hline
\multicolumn{2}{c|}{Hyperparameters} & \multicolumn{3}{c|}{DRIVE}                                                            & \multicolumn{3}{c|}{2018DSB}                                                          & \multicolumn{3}{c}{CVC-ClinicDB}                                                     \\
\hline
\multicolumn{1}{c}{Alpha}                & \multicolumn{1}{c|}{Beta}                & \multicolumn{1}{c}{DSC} & \multicolumn{1}{c}{Precision} & \multicolumn{1}{c|}{Recall} & \multicolumn{1}{c}{DSC} & \multicolumn{1}{c}{Precision} & \multicolumn{1}{c|}{Recall} & \multicolumn{1}{c}{DSC} & \multicolumn{1}{c}{Precision} & \multicolumn{1}{c}{Recall} \\
\hline
1                                        & 0                                       & 0.8082                  & \textbf{0.8473}               & 0.7766                     & 0.9147                  & \textbf{0.9205}               & 0.9168                     & 0.8826                  & 0.9175                        & 0.8759                     \\
0.9                                      & 0.1                                     & \textbf{0.8141}         & 0.8115                        & 0.8211                     & \textbf{0.9171}         & 0.9139                        & 0.9281                     & 0.8928                  & 0.9211                        & 0.8844                     \\
0.8                                      & 0.2                                     & 0.8099                  & 0.7870                        & 0.8386                     & 0.9162                  & 0.9129                        & 0.9268                     & 0.8903                  & 0.9146                        & \textbf{0.8892}            \\
0.7                                      & 0.3                                     & 0.8025                  & 0.7542                        & 0.8621                     & 0.9119                  & 0.8952                        & 0.9374                     & 0.8787                  & 0.8985                        & 0.8811                     \\
0.6                                      & 0.4                                     & 0.7838                  & 0.6759                        & 0.8829                     & 0.9043                  & 0.8873                        & 0.9319                     & \textbf{0.8962}         & \textbf{0.9229}               & 0.8856                     \\
0.5                                      & 0.5                                     & 0.7347                  & 0.6158                        & \textbf{0.9156}            & 0.8742                  & 0.8197                        & \textbf{0.9500}            & 0.8903                  & 0.9143                        & 0.8874    \\\hline                
\end{tabular}}
\label{tab:table2}
\end{table*}

\subsection{Dataset descriptions and evaluation metrics}

To evaluate BU, we select three well-validated, open-source biomedical imaging datasets, namely: Digital Retinal Images for Vessel Extraction (DRIVE), 2018 Data Science Bowl (2018DSB) and CVC-ClinicDB \cite{staal2004ridge, caicedo2019nucleus, bernal2015wm}. The DRIVE dataset contains 40 coloured fundus photographs used for retinal vessel segmentation, 2018DSB consists of 670 light microscopy images for nuclei segmentation, and the CVC-ClinicDB dataset consists of 612 frames of colorectal polyps obtained during optical colonoscopy. A summary of the datasets and training details are presented in Table \ref{tab:table1}. 

\begin{table}[ht!]
\centering
\caption{Details of datasets and training setup used for our experiments.}
\scalebox{0.5}{
\begin{tabular}{lccccccc}
\hline
Dataset      & Segmentation     & \#Images & Size      & \#Training & \#Validation & \#Test \\
\hline
DRIVE        & Retinal vessel   & 40       & $512 \times 512$ & 16         & 4            & 20 \\
2018DSB      & Cell nuclei      & 670      & $256 \times 256$ & 428        & 108          & 134 \\
CVC-ClinicDB & Colorectal polyp & 612      & $288 \times 384$ & 392        & 98           & 122\\ \hline
\end{tabular}}
\label{tab:table1}
\end{table}

For evaluation, we calculate Dice Similarity Coefficient (DSC), precision and recall metrics per image and average over the hold-out test set.

\begin{table*}[t]
\centering
\caption{Evaluating robustness of the Dice loss with and without BU to under-segmentation and over-segmentation. The highest scores are denoted in bold.}
\scalebox{0.75}{
\begin{tabular}{l|l|lll|lll|lll}
\hline
                       &                          & \multicolumn{3}{c|}{DRIVE}                                                            & \multicolumn{3}{c|}{2018DSB}                                                         & \multicolumn{3}{c}{CVC-ClinicDB}                                                     \\
                       \hline
Segmentation           & \multicolumn{1}{c|}{Loss} & \multicolumn{1}{c}{DSC} & \multicolumn{1}{c}{Precision} & \multicolumn{1}{c|}{Recall} & \multicolumn{1}{c}{DSC} & \multicolumn{1}{c}{Precision} & \multicolumn{1}{c|}{Recall} & \multicolumn{1}{c}{DSC} & \multicolumn{1}{c}{Precision} & \multicolumn{1}{c}{Recall} \\
\hline
\multirow{2}{*}{Under} & DSC                      & 0.4387                  & \textbf{0.9845}               & 0.2830                     & 0.8040                  & \textbf{0.9653}               & 0.7034                     & 0.8909                  & \textbf{0.9433}               & 0.8602                     \\
                       & DSC   + BU               & \textbf{0.7198}         & 0.8797                        & \textbf{0.6120}            & \textbf{0.9130}         & 0.9231                        & \textbf{0.9107}            & \textbf{0.8940}         & 0.9225                        & \textbf{0.8857}            \\
                       \hline
\multirow{2}{*}{Over}  & DSC                      & 0.6730                  & 0.5236                        & \textbf{0.9471}            & 0.8417                  & 0.7514                        & \textbf{0.9709}            & 0.8742                  & 0.8773                        & \textbf{0.9006}            \\
                       & DSC   + BU               & \textbf{0.8093}         & \textbf{0.8286}               & 0.7949                     & \textbf{0.9160}         & \textbf{0.9108}               & 0.9291                     & \textbf{0.8797}         & \textbf{0.8975}               & 0.8888        \\ \hline            
\end{tabular}}
\label{tab:table3}
\end{table*}

\subsection{Implementation details}

For our experiments, we used the Medical Image Segmentation with Convolutional Neural Networks (MIScnn) open-source Python library \cite{muller2021miscnn}. All experiments made use of Keras with Tensorflow backend, run on NVIDIA P100 GPUs. For all experiments, except for the DRIVE dataset which is already divided into 20 training images and 20 testing images, we randomly partitioned each dataset into 80\% development and 20\% test set, with further division of the development set into 80\% training set and 20\% validation set. We used the following data augmentation: scaling, rotation, mirroring, elastic deformation and brightness. All images were normalised to $[0,1]$ using the z-score.

Model parameters were initialised using the Xavier initialisation. We trained each model with instance normalisation \cite{zhou2019normalization}, using the Adam optimizer with a batch size of 1 and initial learning rate of $1\times10^{-3}$, and used ReduceLROnPlateau to reduce the learning rate by 0.1 if the validation loss did not improve after 25 epochs. The EarlyStopping callback was used to terminate training if the validation loss did not improve after 50 epochs.

SLs are set to $0.9$ for the segmentation target and $0.1$ for the background \cite{wang2018focal}. For BU, a grid search was performed to select optimal $\alpha$ and $\beta$ values for each dataset. For this preliminary study, A $3 \times 3$ square structuring element was used, and we set $n = 1$. 

To simulate under-segmentation and over-segmentation, we use morphological erosion and dilation operations, respectively, on the manual segmentations in the training set. In these experiments we used the unbalanced form of BU, setting $\alpha = 1$ and $\beta = 1$ for under-segmentation, and $\alpha = 0$ and  $\beta = 0$ for over-segmentation.

\begin{table}[t]
\centering
\caption{Performance comparisons using different ground truth transformations on the DRIVE, 2018DSB and CVC-ClinicDB datasets. The highest scores are denoted in bold.}
\scalebox{0.52}{
\begin{tabular}{l|lll|lll|lll}
\hline
            & \multicolumn{3}{c|}{DRIVE}                                                            & \multicolumn{3}{c|}{2018DSB}                                                         & \multicolumn{3}{c}{CVC-ClinicDB}                                                     \\
            \hline
Loss        & \multicolumn{1}{c}{DSC} & \multicolumn{1}{c}{Precision} & \multicolumn{1}{c|}{Recall} & \multicolumn{1}{c}{DSC} & \multicolumn{1}{c}{Precision} & \multicolumn{1}{c|}{Recall} & \multicolumn{1}{c}{DSC} & \multicolumn{1}{c}{Precision} & \multicolumn{1}{c}{Recall} \\
\hline
DSC         & 0.8082                  & 0.8473                        & 0.7766                     & 0.9147                  & 0.9205                        & 0.9168                     & 0.8826                  & 0.9175                        & 0.8759                     \\
DSC   + SL  & 0.8032                  & 0.7420                        & \textbf{0.8797}            & 0.8863                  & 0.8478                        & \textbf{0.9531}            & 0.8660                  & 0.8417                        & \textbf{0.9267}            \\
DSC   + DPT & 0.8086                  & \textbf{0.8498}               & 0.7751                     & 0.9016         & \textbf{0.9230}               & 0.8891                     & 0.8174                  & 0.9058                        & 0.7614                     \\
DSC   + BU  & \textbf{0.8141}         & 0.8115                        & 0.8211            & \textbf{0.9171}                  & 0.9139                        & 0.9281                     & \textbf{0.8962}         & \textbf{0.9229}               & 0.8856   \\ \hline                 
\end{tabular}}
\label{tab:table4}
\end{table}

\section{Results}

\subsection{Hyperparameter tuning}

The results for the hyperparameter tuning experiments are shown in Table \ref{tab:table2}. The highest DSC values were obtained with $\alpha = 0.9, \beta = 0.1$ for the DRIVE and 2018DSB experiments with a score of 0.8141 and 0.9171 respectively, higher than 0.8082 and 0.9147 obtained without BU. For CVC-ClinicDB, the highest DSC values of 0.8962 was obtained with $\alpha=0.6, \beta = 0.4$, compared to 0.8826 with the original DSC loss. The lower optimal $\alpha$ value for the CVC-ClinicDB dataset correlates with the greater uncertainty associated with delineating colorectal polyp boundaries compared to retinal vessels and cell nuclei. The general increase in recall and decrease in precision with higher values of $\alpha$, despite using the balanced form, is a property of using the Dice loss, which is biased towards assignment of the under-represented class.

\subsection{Robustness to systematic segmentation error}

The results for evaluating robustness to under-segmentation and over-segmentation are shown in Table \ref{tab:table3}. Including BU is associated with the highest DSC values across datasets for both under-segmentation and over-segmentation. The difference was most apparent in the DRIVE dataset, where the DSC score reduced to 0.4387 with under-segmented ground truth labels without BU, while only reduced to 0.7198 with BU. The smallest differences were observed with the CVC-ClinicDB, which as previous mentioned, may be due to the greater uncertainty associated with demarcating polyp boundaries.

\subsection{Performance comparisons using different ground truth transformations}

The results for different ground truth transformations are shown in Table \ref{tab:table4}. The highest DSC scores were consistently observed with BU. Using SL and DPT generally negatively affected the DSC score. By applying SL globally, the label noise generated by regions distant to the boundary may prevent optimal convergence. In contrast, DPT may attend to boundary regions too strictly, resulting in overfitting.

\begin{figure}[!ht]
\centering
    \includegraphics[width=0.45\textwidth]{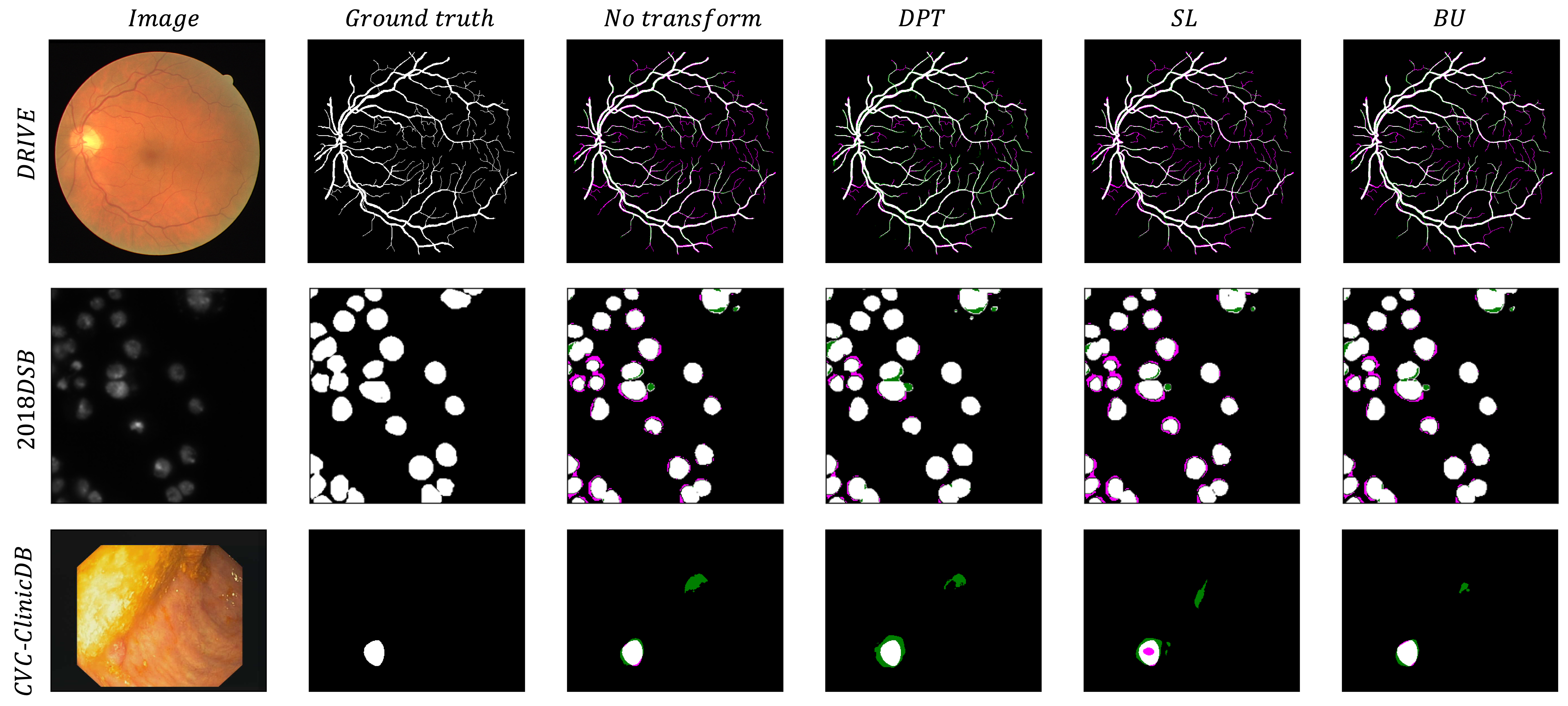}
    \caption{Example segmentations using each ground truth transformation for the three datasets. False positive and false negative predictions are highlighted in green and purple, respectively.}
    \label{fig:figure_3}
\end{figure}

Example segmentations are shown in Fig.~\ref{fig:figure_3}. Segmentations produced using BU are the most similar to the ground truth, even with difficult segmentations such as the CVC-ClinicDB example.

\begin{figure}[!ht]
\centering
    \includegraphics[scale=0.35]{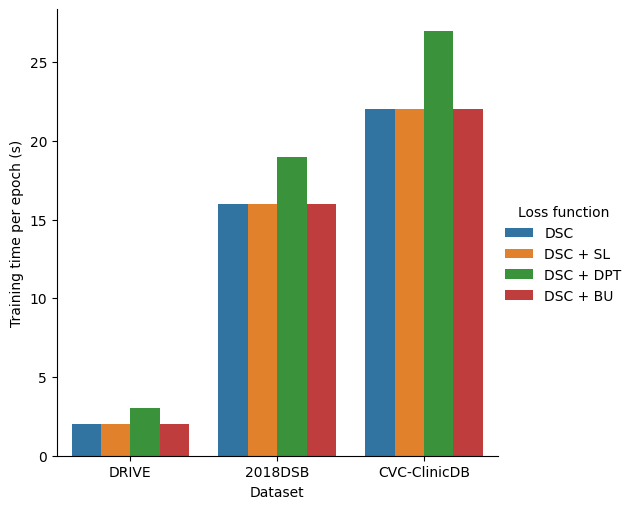}
    \caption{Efficiency comparisons of different ground truth transformations on the DRIVE, 2018DSB and CVC-ClinicDB datasets.}
    \label{fig:figure_4}
\end{figure}


\subsection{Efficiency comparisons of different ground truth transformations}

Although ground truth transformations may be pre-computed prior to training, this is not possible when on-the-fly data augmentation is used. The training times using different ground truth transformations are shown in Fig.~\ref{fig:figure_4}. Only DPT was associated with an increase in training time, demonstrating the efficiency of BU.

\section{Conclusion}

In this work, we developed a new ground truth transformation, known as BU, to reflect segmentation uncertainty, by using morphological operations to restrict SL to boundary regions. We demonstrated improved performance using BU over global SL and DPT, across three well-validated biomedical imaging datasets. Moreover, we showed how BU may be adapted to provide robustness to over and under-segmentation. Finally, we confirmed that BU is an efficient operation, with no increase in training time. Appreciating the underlying uncertainty with manual segmentations not only more accurately reflects the nature of manual segmentation with biomedical images, but may also improve model generalisation to unseen datasets by reducing overfitting.

\clearpage
\section{Compliance with Ethical Standards}
This research study was conducted retrospectively using open-source medical imaging datasets. Ethical approval was not required as confirmed by the license attached with the open access data.

\bibliographystyle{IEEEbib}
\bibliography{strings,refs}

\end{document}